\documentclass[prl,twocolumn,superscriptaddress,nobibnotes]{revtex4}
\usepackage{graphicx}
\usepackage{amssymb}
\usepackage{amsmath}
\usepackage{bm}
\usepackage{color}
\usepackage{float}
\usepackage{setspace}
\usepackage{physics}

\usepackage{subfigure}

\newcommand{\q}{\mathbf{q}}

\newcommand{\nxsigma}{n^{\operatorname{X}}_\sigma}

\begin{document}

\title{Exciton interaction induced spin splitting in MoS$_2$ monolayer}

\author{Yao Li}
\affiliation{Tianjin Key Laboratory of Molecular Optoelectronic Science, Institute of Molecular Plus, School of Science, Tianjin University, Tianjin 300072. China} 

\author{G. Li}
\email[]{guangyao.li@monash.edu}
\affiliation{School of Physics and Astronomy, Monash University, Victoria 3800, Australia}
\affiliation{ARC Centre of Excellence in Future Low-Energy Electronics Technologies, Monash University, Victoria 3800, Australia}

\author{Xiaokun Zhai}
\affiliation{Institute of Functional Crystals, Tianjin University of Technology, Tianjin 300384, China}

\author{Shi Fu Xiong}
\affiliation{Institute of Functional Crystals, Tianjin University of Technology, Tianjin 300384, China}

\author{Hongjun Liu}
\affiliation{Institute of Functional Crystals, Tianjin University of Technology, Tianjin 300384, China}

\author{Xiao Wang}
\affiliation{Key Laboratory for Micro-Nano Physics and Technology of Hunan Province, School of Physics and Electronics, Hunan University, Changsha, Hunan, 410012, China}

\author{Haitao Chen}
\affiliation{College of Advanced Interdisciplinary Studies, National University of Defence Technology, Changsha, 410073, Hunan, China}

\author{Ying Gao}
\affiliation{Tianjin Key Laboratory of Molecular Optoelectronic Science, Institute of Molecular Plus, School of Science, Tianjin University, Tianjin 300072. China} 

\author{Xiu Zhang}
\affiliation{Tianjin Key Laboratory of Molecular Optoelectronic Science, Institute of Molecular Plus, School of Science, Tianjin University, Tianjin 300072. China} 

\author{Tong Liu}
\affiliation{Department of Aerospace Science and Technology, Space Engineering University, Beijing 101416, China} 

\author{Yuan Ren}
\affiliation{Department of Aerospace Science and Technology, Space Engineering University, Beijing 101416, China}

\author{Xuekai Ma}
\affiliation{Department of Physics and Center for Optoelectronics and Photonics Paderborn (CeOPP), Universit\"{a}t Paderborn, Warburger Strasse 100, 33098 Paderborn, Germany}

\author{Hongbin Fu}
\affiliation{Tianjin Key Laboratory of Molecular Optoelectronic Science, Institute of Molecular Plus, School of Science, Tianjin University, Tianjin 300072. China} 

\author{T. Gao}
\email[]{tinggegao@tju.edu.cn}
\affiliation{Tianjin Key Laboratory of Molecular Optoelectronic Science, Institute of Molecular Plus, School of Science, Tianjin University, Tianjin 300072. China} 

\begin{abstract}
{By pumping nonresonantly a MoS$_2$ monolayer at $13$ K under a circularly polarized cw laser, we observe exciton energy redshifts that break the degeneracy between B excitons with opposite spin. The energy splitting increases monotonically with the laser power reaching as much as $18$ meV, while it diminishes with the temperature. The phenomenon can be explained theoretically by considering simultaneously the bandgap renormalization which gives rise to the redshift and exciton-exciton Coulomb exchange interaction which is responsible for the spin-dependent splitting. Our results offer a simple scheme to control the valley degree of freedom in MoS$_2$ monolayer and provide an accessible method in investigating many-body exciton exciton interaction in such materials.}
\end{abstract}

\maketitle

Transition metal dichalcogenides (TMDs) attracted great attention recently because of their unique photonic and optoelectronic properties %.characteristics and their potentials for photonic and electronic devices 
\cite{2D review_xu and heinz, 2D review_Coleman, WangRMP2018}.
When a bulk TMD is thinned down to a monolayer (ML), it becomes a direct bandgap material with gaps located at $K$ and $K^\prime$ points \cite{SplendianiNanoLett2010,MakPRL2010} in the corners of the Brillouin zone. Owning to the spin-orbit interactions, each valley is spin-split, resulting in two distinct exciton resonances denoted as A and B excitons \cite{MakPRL2010,MuellerNpj2018}. TMD excitons' binding energy (about $0.5$ eV) is much larger than that of GaAs ($10$ meV), with a smaller Bohr radius $a_B$ ($1$ nm \emph{vs.} $10$ nm), so that their excitonic features are robust even at room temperature \cite{WangRMP2018}, and providing a solid platform to investigate  exciton physics over a broad range of temperature and excitation density \cite{bandgap renormalization_heinz, bandgap renormalization_Xu, blueshift and redshift}.
Since the states in the $K$ and $K^\prime$ valley are related with each other by time-reversal symmetry, the corresponding exciton energies of the same type from different valley are degenerate \cite{SplendianiNanoLett2010,valley spin _Yao,LiuPRB2013}. In the interest of engineering TMD based valleytronic devices, it is desirable to lift the energy degeneracy to make use of the valley degree of freedom \cite{SchaibleyNatRev2016}. In experiments, it has been demonstrated that the exciton degeneracy can be broken by an out-of-plane magnetic field up to $10$ T \cite{LiPRL2014,magnetic control-Ima,MacNeillPRL2015,AivazianNatPhys2015,enhance energy splitting-Tingyu,magnetic substrate_Zeng,WS2 magnetic promimity effect}. Such apparatus, however, might not be applicable in the realistic applications. 

Another experimentally accessible method to lift the degeneracy is employing the optical Stark effect \cite{Gedik optical stark effect,Wang Feng optical stark effect}. When a pulsed laser pump the TMD ML with energy just below the exciton resonance, the hybridization between the equilibrium state and the photon-dressed state will shift the exciton to a larger energy. By applying a pulsed probe laser to measure the change of absorption/reflection spectra, the exciton energy degeneracy between different valleys is lifted if the pump laser is circularly polarized \cite{Gedik optical stark effect,Wang Feng optical stark effect}, which is supposedly equivalent to a $60$ T pseudomagnetic field for a $10$ meV splitting. This approach, however, not only requires high laser power (GW/cm$^2$) but also it is intrinsically transient, which can only reveal a TMD ML's dynamical properties within the pump pulse duration. Therefore, an optically-tunable exciton splitting scheme applicable to the equilibrium scenario in the TMD ML is still missing.

In this work, we demonstrate experimentally that the energy degeneracy of excitons of the same type of different valley in a MoS$_2$ ML can be broken by the spin-dependent exciton-exciton interactions via a circularly polarized cw pump at $13$ K. The laser energy is 230meV larger than the exciton resonance in MoS$_2$ ML. The spin splitting can be tuned from 0 to $18$ meV under the pump power of $0.3$ MW/cm$^2$. Theoretically, the splitting results from the disagreement of exciton-exciton interaction between different valley which raises the exciton energies differently \cite{FernandezPRB1996}, on top of an overall band gap renormalization given by free carries \cite{HaugPQE1985,EllJOSAB1989}. Our results offer a new way to control the valley degree of freedom, and could be directly applicable to TMD ML exciton-polariton systems \cite{WangNanoLett2016,LiuNatPho2015}.

\emph{Experiment}. 
A piece of MoS$_2$ ML is grown at $760$ K by the chemical vapor deposition (CVD) on a SiO$_2$/Si substrate. %where the thickness of the  SiO$_2$ layer is around $300$ nm. 
To protect the ML, a hBN layer with thickness of around $10$ nm is transferred on top of the sample, shown in the atomic force microscopy (AFM) characterization in Fig.~\ref{fig:sample}(a). We measure the optical spectrum of the MoS$_2$ ML by a home-built confocal PL setup (see Supplemental Material for more details) and obtain smooth PL spectra originated from the recombination of A excitons ($1.865$ eV) and B excitons ($2.024$ eV), as demonstrated in Fig.~\ref{fig:sample}(b). The energy difference between A and B excitons of about 0.159 eV agrees with previous measurements \cite{MoS2-Marie, Tony heinz-MoS2, A-B exciton energy difference, valley relaxation_Finazzi}. Fig.~\ref{fig:sample}(c) shows the Raman spectrum measured at room temperature by a commercial Raman microscope (WITec alfa 300), and the peak at $384.67$ cm$^{-1}$ confirms the good quality of the MoS$_2$ ML. 

\begin{figure}[t]
	\centering
	\includegraphics[width=8.5 cm]{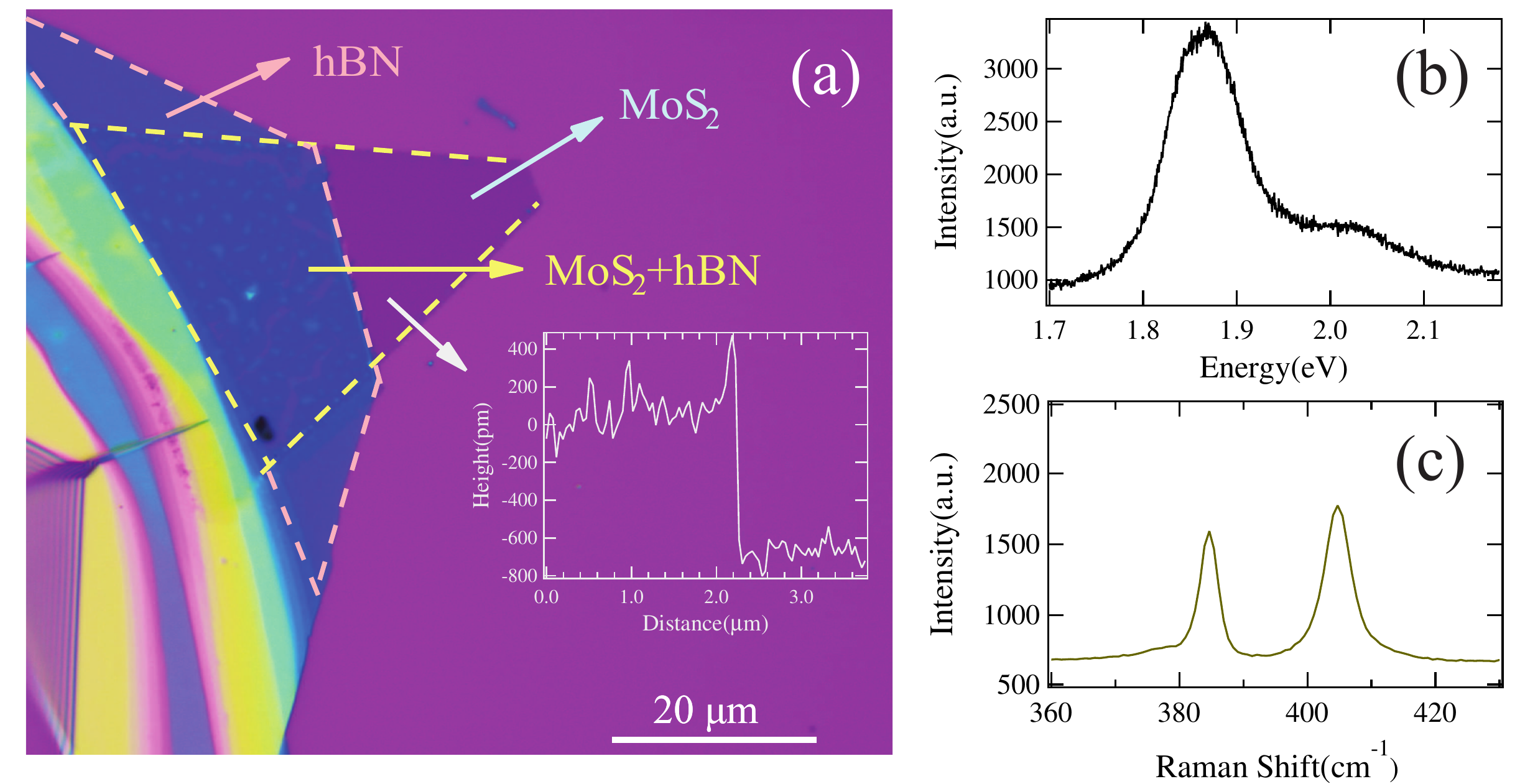} 
	\caption{Optical characterization of the sample. (a) AFM mapping of the MoS$_2$ ML, where the inset shows the thickness of the sample is about $0.7$ nm. (b) PL spectrum of the MoS$_2$ ML excited by a $532$ nm laser at room temperature. (c) Raman spectrum of the sample at room temperature.} \label{fig:sample}
\end{figure}

To study exciton-exciton interactions, the sample is cooled down to $13$ K by a cryogen free cryostat (Janis). We use a $532$ nm laser (Spectra Physics) as the excitation source. The laser can be adjusted from linear to circular (left-handed or right-handed) polarization by using a quarter waveplate. The laser is focused on the sample through an objective with a spot of around 2 $\mu$m , and the emitted photons are collected by the same objective. A quarter waveplate, half waveplate and a linear polarizer are combined to resolve the polarization of the emitted PL from the sample. The polarization of photons is calculated by $P=(I^{+}-I^{-})/(I^{+}+I^{-})$, where $I^{+}$ and $I^{-}$ denote the counts of the right-hand and left-handed circularly polarization components.

\begin{figure}[t]
	\centering
	\includegraphics[width=8.5 cm]{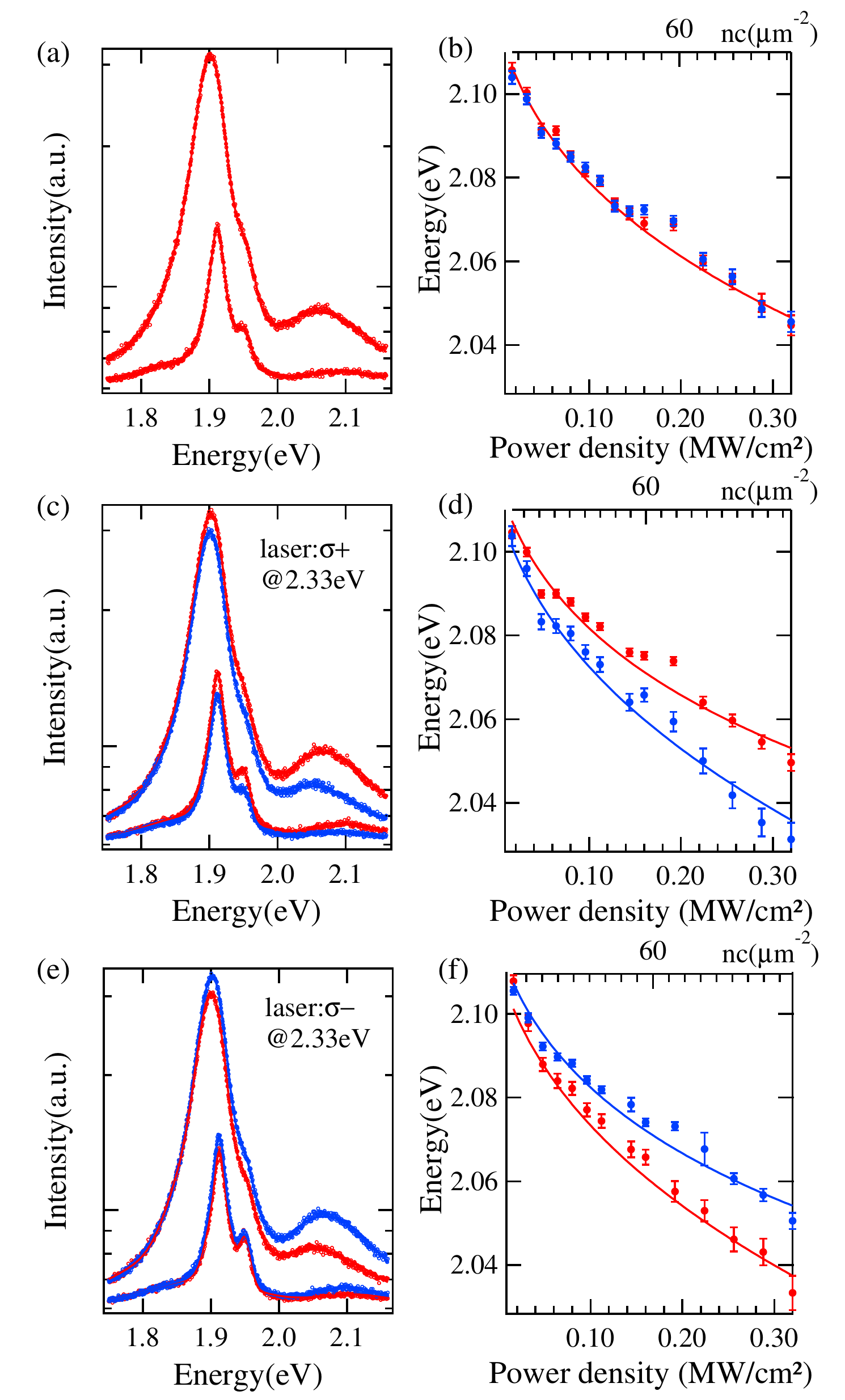}
	\caption{The spectrum of the MoS$_2$ monolayer as a function of the pumping power. (a)(c)(e) Optical spectrum of the sample pumping by a linearly polarized, right-handed circularly polarized, and left-handed circularly polarized laser respectively, with density of 0.03MW/cm$^2$ and 0.3MW/cm$^2$. (b)(d)(f) PL peaks (fitted B exciton energy) against the pumping power density (bottom axis), with polarization matching the corresponding spectra on the left columns. Red (blue) dots correspond to B (B$^\prime$) excitons. Solid curves are the theoretical plots against $n_c$ (top axis) according to Eq.~\eqref{eq:PL peak expression}. Theory parameters can be found in \cite{theoryparameters}. } \label{fig:red shift}
\end{figure}

At $13$ K, the energies of A and B excitons are blueshifted to $1.948$ eV and $2.099$ eV respectively, and the trion peak appears at $1.911$ eV \cite{ChristopherSciRep2017} under the pump density of 0.008MW/cm$^2$, see Fig.~\ref{fig:red shift}(a). During the experiment, a mechanical chopper with a duty circle of $5\%$ is used to reduce possible heating effect by the laser. When the pump power is continuously increased, the spectrum peaks corresponding to exciton energies shift to lower values, i.e. redshifted, see Fig.~\ref{fig:red shift}(b).   %The redshift of exciton energy can be explained by the bandgap renormalization and plasmon screening by the free carriers \cite{blueshift and redshift}. 
Similar redshift of the exciton energy has been reported previously in \cite{blueshift and redshift}, where the anomalous blueshift and redshift crossover was measured by a pump-probe spectroscopy with a WS$_2$ ML and it was interpreted phenomenological by a Lennard-Jones potential between excitons \cite{blueshift and redshift}. In our current experiment, we did not observe the redshift-blueshift crossover.

Next we discuss the exciton energy changes with the spin degree of freedom. Since the A exciton peak overlaps with the trion peak \cite{ChristopherSciRep2017} and it was demonstrated that the valley polarization effect of B excitons is stronger than that of A excitons \cite{MoS2-Marie}, here we focus on B excitons for the spin-split effect and sometimes refer them simply as excitons. Figure~\ref{fig:red shift}(c) shows the PL spectra of the MoS$_2$ ML pumping by a right-hand circularly polarized laser. The non-zero left-hand polarized PL is presumably caused by intervalley scattering \cite{Wang Feng optical stark effect}, and the polarization $P$ of B excitons is about $30\%$ in agreement with \cite{MoS2-Marie}. By examining the spectral peak positions of each polarization component, we find an energy splitting about $18$ meV under the pump power of 0.3MW/cm$^2$. The energy splitting is reversed when the pump switches to the opposite polarization, see Fig.~\ref{fig:red shift}(e). In both cases, the energy of the co-polarized component is always higher than that of the cross-polarized component, in contract to the results of MoSe$_2$/WSe$_2$ heterostructures \cite{Gao spin splitting}. When the pump laser is linearly polarized, the energy splitting between two circularly polarized components disappears, see Fig.~\ref{fig:red shift}(a)(b).

The spin dependent energy splitting can be explained by the exciton-exciton interaction originated from the Coulomb interaction between electrons and holes. Simply speaking, the population imbalance between B excitons in different valley gives rise to a different energy correction because of the Coulomb interaction. Similar splitting can also be found in conventional semiconductor quantum well systems such as GaAs or InAs \cite{Luis vina spin splitting PRL, Luis vina spin splitting,FernandezPRB1996,X.Marie spin exciton exciton interaction,InAs spin splitting}. However, the spin splitting observed is created under pulsed excitation and greatly reduced %(1meV)
when the laser is tuned far away from the exciton resonance \cite{InAs spin splitting}. Thanks to the spin valley locking mechanism, the laser energy to induce the exciton splitting can be much larger (230meV) than the exciton resonance in MoS$_2$ ML, this leaves greater flexibility for the laser choice to control the valley degrees of freedom. 

%Our result is different from the works in III-IV quantum well and bulk semiconductors \cite{Luis vina spin splitting PRL, Luis vina spin splitting,FernandezPRB1996,X.Marie spin exciton exciton interaction,InAs spin splitting}, where the spin splitting is created under pulsed excitation and greatly reduced %(1meV)  when the laser is tuned far away from the exciton resonance \cite{InAs spin splitting}. 

Compared with the optical Stark effect \cite{Gedik optical stark effect,Wang Feng optical stark effect} in the TMDs ML, the exciton splitting can be manipulated under much lower CW pumping density. 
%our pump laser is cw and it is far blue-detuned from the exciton resonance (the detuning will increase alongside of the pump power, as the effective band gap gets renormalized). 
In supplemental materials, we plot the exciton splitting as a function of the pumping density.  The exciton splitting reaches 18 meV at the pumping density of around 0.3MW/cm$^2$, around four orders of magnitude smaller than the optical Stark effect \cite{Wang Feng optical stark effect}. In addition, we can tune the exciton splitting from zero to 18meV continuously, which cannot be realized in \cite{Gao spin splitting}. Thus, the interexcitonic exchange interaction offers to tune the exciton splitting more effectively and stably in MoS$_2$ ML.In the following we develop a theory to model the energy splitting of the MoS$_2$ ML under various pumping configurations.

\emph{Theory.} The measured exciton PL peak energy $E_{PL}$ corresponds to recombination of electrons in the conduction band with the bound holes in valence bands, which can be written as $E_{PL}=E_g+(-E_B)$, where $E_g$ is the bandgap and $-E_B$ is the exciton binding energy. With increased carrier density, it has been shown that the excessive free carriers will result in bandgap renormalization and screening of $-E_B$ \cite{blueshift and redshift}. First we look at the bandgap renormalization. After taking the random phase approximation and quasi-static approximation, the energy reduction of the bandgap can be expressed as \cite{HaugPQE1985,EllJOSAB1989}:
\begin{equation}\label{eq:bandgap_renor}
\Delta E_g (n_c)=V_s (k_F)\, n_c+\sum_\q\left[V_s(\q)-V(\q)\right],
\end{equation}
where $n_c$ is the free carrier density (assuming $n_e=n_h=n_c$), $k_F=(2\pi n_c)^{1/2}$ is the $2D$ Fermi wave number, $V(\q)=e^2/(2\epsilon_0 q)$ is the unscreened $2D$ Coulomb potential in SI unit with $\epsilon_0$ the background permittivity, and $V_s(\q,\omega)=V(\q)/\varepsilon(q,\omega)$ is the screened Coulomb potential. The screening effect can be simplified by the plasmon-pole approximation: $1/\varepsilon=1+\omega^2_{pl}/[(\omega+i\delta)^2-\omega^2_q]$, where $\omega_{pl}(q)=[2\pi e^2 n_c q/(\epsilon_0 m_r)]^{1/2}$ is the plasma frequency with $m_r$ the reduced electron-hole mass, $\omega$ is the light frequency, $\delta$ is a small number that shifts the pole away from the real axis, $\omega^2_q=\omega^2_{pl}(1+q/\kappa)+\nu^2_q$, and $\nu^2_q=C[\hbar q^2/(4 m_r)]^2$; $\kappa=\frac{2\pi e^2}{\epsilon_0}\sum_{e,h}\frac{\partial n_c}{\partial \mu_{e,h}}$ is the screening wave number, with $\mu_{e,h}$ and $n_c$ related by the Fermi function \cite{EllJOSAB1989}.

Next we look at screening and spin imbalance effects on the exciton energy. The exciton binding energy will reduce to $-E_B/\varepsilon_r^2$ because of screening \cite{excitonbook}, with $\varepsilon_r$ representing the increased dielectric constant.
In the presence of valley degree of freedom, if we ignore the Coulomb exchange between excitons from different valleys (equivalent to setting the excitons' center-of-mass momentum to zero \cite{YuNatComm2014}), the meanfield first order exciton energy correction is: $\Delta E_\sigma\propto E_B \nxsigma$ \cite{FernandezPRB1996}, where $\nxsigma$ is the excition density and $\sigma=\pm$ is the pump beam polarization (or exciton valley) index. Under an incoherent cw pump, the values of $\nxsigma$ and $n_c$ are not directly measurable.
Considering that in forming an exciton one need to choose independently an electron and a hole, and that $n_e=n_h=n_c$, we can assume $\nxsigma\sim n_c^2$ so that $\Delta E_\sigma\sim n_c^2$. Mathematically, when $n_c$ is small (in units of $a_B^2$), the Maclaurin series of $\nxsigma$ reads: $\nxsigma\approx a_\sigma+b_\sigma\, n_c +c_\sigma\, n_c^2$ with $a_\sigma,b_\sigma,c_\sigma$ the expansion coefficients. It can be rewritten in vertex form $\nxsigma=a_\sigma^\prime+b_\sigma^\prime (n_c -c_\sigma^\prime)^2$. We redefine the shifted $n_c -c_\sigma^\prime$ as $n_c$ \cite{theoryexplain} and treat $b_\sigma^\prime$ as a fitting parameter denoted as $f_\sigma$. Finally, the sum of two $n_c$-independent terms $a_\sigma^\prime$ and $-1/\varepsilon_r^2$ is denoted as $s_\sigma$, and we reach the screened exciton energy:
\begin{equation}\label{eq:exciton energy}
\Delta E^{\operatorname{X}}_\sigma (n_c)=(s_\sigma+f_\sigma\,n_c^2)E_B.
\end{equation}
Combining Eqs.~\eqref{eq:bandgap_renor} and \eqref{eq:exciton energy}, we can express the PL peak energy for excitons from different valleys as:
\begin{equation}\label{eq:PL peak expression}
E_\sigma (n_c)=E_g+\Delta E_g(n_c)+(s_\sigma+f_\sigma\,n_c^2)E_B,
\end{equation} 
where $s_\sigma$ and $f_\sigma$ are fitting parameters corresponding to the spin imbalance of the pump. Fig.~\ref{fig:red shift} shows the theoretical curves  %Eq.~\eqref{eq:PL peak expression} 
plotting along with experimental data, from where we can estimate the free carrier density $n_c\sim 10^9\,-\,10^{10}$ cm$^{-2}$ similar to that observed in a previous experiment \cite{blueshift and redshift}. Note that if we extend the theoretical plot of Eq.~\eqref{eq:PL peak expression} beyond the current experimental range, the PL peak will exhibit blueshift behavior when the energy correction term $f_\sigma\,n_c^2$ surpasses the other terms. 
In this case, the meanfield first order perturbation theory breaks down and a more rigorous calculation would be required. Nevertheless, based on the existing experimental data we can see that the nonlinear dependence of $n_X$ on $n_c$ could be responsible for the redshift-blueshift crossover, which was previously explained by the Lennard-Jones type potential in real space \cite{blueshift and redshift}. 
 
Our experimental results can be repeated in other MoS$_2$ samples using different fabrication techniques. In the supplemental materials, we show the existence of the spin splitting in a CVD grown MoS$_2$ monolayer without hBN capping and a mechanically exfoliated MoS$_2$ monolayer. Although the exciton energy in there samples vary piece by piece, however, the spin splitting does not change too much under the same pumping condition. This roles out the possibility that the spin splitting results from the interaction with the capping hBN layer or some defects in the growth process. In addition, the spin splitting also exists in a MoS$_2$ monolayer using a pure silicon and hBN substrate (shown in the supplemental materials), excluding the effect from the substrate.  

\begin{figure}[t]
	\centering
	\includegraphics[width=8.5 cm]{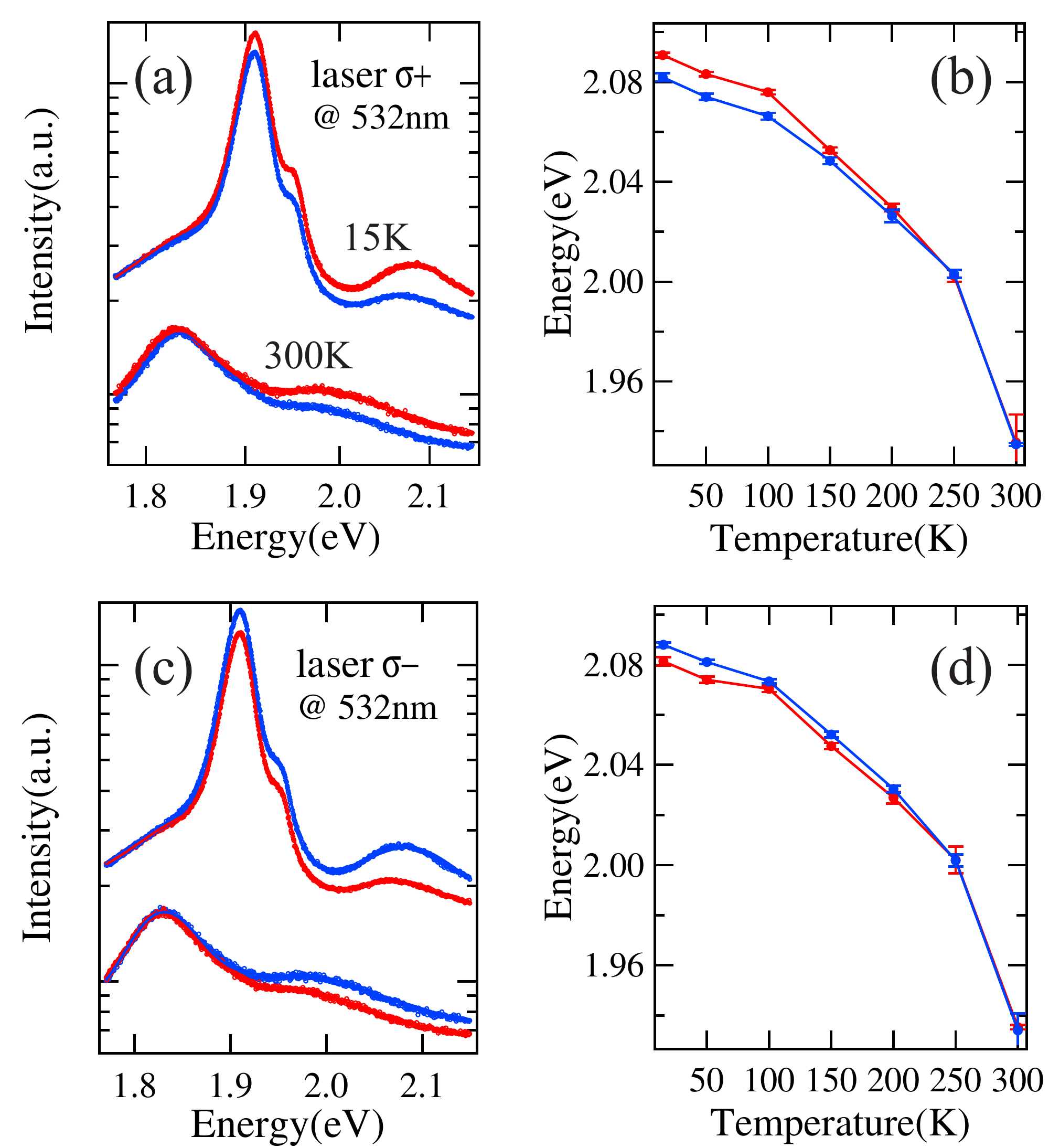} 
	\caption{PL spectra of the MoS$_2$ ML at different temperatures. (a)(c) Optical spectra at $15$ K and $300$ K pumping by a right- and left-handed circularly polarized laser respectively. Power density: 0.03MW/cm$^2$. (b)(d) Exciton energy against temperature.} \label{fig: T dependence}
\end{figure}

With increasing the temperature, both the excitons A and B will shift to lower energies. In Fig3, we repeat the measurements at different temperatures (15K, 50K, 100K, 150K,  200K,  250K and 300K). At higher temperature, the circular polarization of the photoluminescence decreases due to the phonon assisted intervalley scattering, in agreement of \cite{cui valley polarization, MoS2-room temperature circular polarization, near unity CP MoS2, MoS2-Marie}. 
%The intervalley scattering \cite{MoS2-spin relaxation}, e.g. the intraexcitonic exchange interaction can also suppress the valley polarization.  
Although the spectra corresponding to the exciton B can keep the circular polarization even at room temperature (10\%), The spin splitting under the circularly polarized pumping approaches zero at around 200K, as shown in Fig3 (b)(d).  

\emph{Conclusion}. 
We show experimentally and theoretically that spin dependent exciton exciton interaction under a spin-biased pump can lead to exciton energy splitting between different valleys in a MoS$_2$ ML. The spin splitting can be tuned at much lower power density (MW/cm$^2$) approaching 18meV, which is much more effective compared with the optical Stark effect.  In addition, it does not need external magnetic field or stacking different materials. The CW-pumping created spin splitting provides a reliable way to explore the valley degree of freedom for MoS$_2$ ML and it is readily applicable to the investigation of the spin dynamics of TMD exciton-polariton systems.

%The interexcitonic exchange interactions can manipulate the pseudospin of excitons and modifies the exciton's energy distribution with different spins under moderate pump power,  which have important implications on spin related many body interactions and spin optoelectronic devices based on the TMD monolayers.  

\begin{acknowledgments}

TG acknowledges supports from National Natural Science Foundation of China (NSFC, No. 11874278) and supports from Young scholar 1000 Talents plan.
GL thanks Dr. Dmitry Efimkin for useful discussions.

\end{acknowledgments}

%Figures of supplementary material

\appendix

%BiBtex reference style
%\bibliographystyle{vortex_gen_ref_format}
%\bibliography{vortex_gen_reference}

\begin{thebibliography}{99}
\bibitem{theoryexplain} If $\nxsigma$ is expanded only up to a linear term in $n_c$; or in the quadratic expansion the linear term is kept, then in both cases the fitted coefficient of the highest order term would be negative for some data. Physically it means that the exciton density decreases with the increase of free carrier density, which is unreasonable. Therefore, the fitting expression of Eq.~\eqref{eq:exciton energy} is quadratic without a linear term.
\end{thebibliography}

%plain tex reference style

\end{document}